\documentclass[10pt,a4paper,amsmath,amssymb,aps,prl,nofootinbib,twocolumn,nobibnotes]{revtex4-2}
\usepackage{float}
\usepackage{lmodern}
\usepackage{booktabs}
\usepackage[T1]{fontenc}
\usepackage[utf8]{inputenc}
\usepackage{graphicx}
\usepackage{dcolumn}
\usepackage{bm}
\usepackage{color}
\usepackage[normalem]{ulem}

\usepackage[unicode=true,
 bookmarks=true,bookmarksnumbered=false,bookmarksopen=false,
 breaklinks=false,pdfborder={0 0 1},backref=false,colorlinks=true]
 {hyperref}

\definecolor{urlblue}{rgb}{0,0,0.9}\definecolor{linkblue}{rgb}{0,0,.8}\definecolor{linkgreen}{rgb}{0,0.45,0}\definecolor{linkpurple}{rgb}{0.7,0.0,0.4}\definecolor{linkorange}{rgb}{0.7,0.1,0.0}\AtBeginDocument{\hypersetup{
linkcolor=linkblue,
citecolor=linkorange,
urlcolor=urlblue}}\definecolor{urlblue}{rgb}{0,0,0.9}\definecolor{linkblue}{rgb}{0,0,.8}\definecolor{linkgreen}{rgb}{0,0.45,0}\definecolor{linkpurple}{rgb}{0.7,0.0,0.4}\definecolor{linkorange}{rgb}{0.7,0.1,0.0}
\AtBeginDocument{\hypersetup{
linkcolor=linkblue,
citecolor=linkorange,
urlcolor=urlblue}}

\definecolor{commentblue}{rgb}{0,0,.85}

\usepackage[mathlines]{lineno}

\newcommand{\betad}{\boldsymbol{\beta}^{\rm D}}
\newcommand{\betaa}{\boldsymbol{\beta}^{\rm A}}
\newcommand{\dint}{\boldsymbol{\Delta}_{1,{\rm int}}}
\newcommand{\done}{\boldsymbol{\Delta}_{1}}

\newcommand\T{\rule{0pt}{2.6ex}}       

\graphicspath{{prl-plots/}}

\begin{document}

\title{First constraints on the intrinsic CMB dipole and our velocity \\ with Doppler and aberration}

\author{Pedro da Silveira Ferreira}
\thanks{Both authors contributed equally}
\affiliation{Observatório do Valongo, Universidade Federal do Rio de Janeiro, 20080-090, Rio de Janeiro, RJ, Brazil}
\author{Miguel Quartin}
\thanks{Both authors contributed equally}
\affiliation{Observatório do Valongo, Universidade Federal do Rio de Janeiro, 20080-090, Rio de Janeiro, RJ, Brazil}
\affiliation{Instituto de Física, Universidade Federal do Rio de Janeiro, 21941-972, Rio de Janeiro, RJ, Brazil}

\date{\today}

\begin{abstract}
    We test the usual hypothesis that the Cosmic Microwave Background (CMB) dipole, its largest anisotropy, is due to our peculiar velocity with respect to the Hubble flow by measuring independently the Doppler and aberration effects on the CMB using Planck 2018 data. We remove the spurious contributions from the conversion of intensity into temperature and arrive at measurements which are independent from the CMB dipole itself for both temperature and polarization maps and both SMICA and NILC component-separation methods. Combining these new measurements with the dipole one we get the first constraints on the intrinsic CMB dipole. Assuming a standard dipolar lensing contribution we can put an upper limit on the intrinsic amplitude: $3.7$mK (95\%~CI). We estimate the peculiar velocity of the solar system without assuming a negligible intrinsic dipole contribution: $v = (300^{+111}_{-93})$ km/s with $(l,b) =(276\pm 33, \,51\pm 19)^\circ$ [SMICA], and  $v~=~(296^{+111}_{-88})$~km/s with $(l,b) =(280\pm 33, \,50\pm 20)^\circ$ [NILC] with negligible systematic contributions. These values are consistent with the peculiar velocity hypothesis of the dipole.
\end{abstract}

\maketitle

\textbf{\emph{Introduction.}}
The CMB temperature dipole is the largest CMB anisotropy not coming from foregrounds, and has been precisely measured since a few decades ago. After removal of the orbital contribution due to the motion of the instrument with respect to the Sun, we are left with the so-called solar dipole, with an amplitude of $3.36208 \pm 0.00099$ mK~\citep{Akrami:2018vks}, which is $\sim100$ larger than the other multipoles. It is thus fully credited to the proper motion between the solar system and the CMB rest frame. If one assumes that the whole dipole has such a kinematic origin, one infers a relative velocity of $(369.82 \pm 0.11)$~km/s. This velocity estimate is widely reported and often used in astronomy in order to convert observed redshifts into CMB-centric (cosmological) ones.

Nevertheless there exists the possibility that part of the dipolar effect could be due to primordial fluctuations in the surface of last-scattering (SLS).
Concrete alternatives to the kinematic scenario were discussed as far back as 30 years ago by~\cite{Paczynski1990}, which showed that a large local void could also explain the dipole. This particular scenario was further investigated by e.g.~\cite{Tomita:1999rw} and~\cite{Quartin:2009xr}. A ``tilted universe scenario'' composed of a superhorizon isocurvature perturbation was proposed in~\cite{Turner_1992}. An inflationary model which produces similar results was proposed by~\cite{Langlois:1996ms}. This scenario can be tested as it leads to a cosmic ``bulk flow'' of galaxies and clusters~\cite{Ma_2011}. In the kinematic dipole hypothesis the $\Lambda$CDM model predicts how the average peculiar motion of a sphere of galaxies should approach zero as the radius increases. Low redshift surveys have attempted to measure this convergence in the last 30 years~\cite{Strauss1992,Dale:1998jt,Maller:2003en,Erdogdu:2005wi,Bilicki:2011wq,Nusser:2014sha}. These measurements have a number of potential systematics~\cite{Carrick:2015xza}, and the reported results have been inconsistent, as discussed by e.g.~\cite{Hoffman:2015waa}.
High-redshift radio galaxies have also been used to test the kinematic hypothesis. Their measured dipole is a few times higher than what would be expected from the CMB dipole~\citep{Colin:2017juj,Bengaly:2017slg,Siewert:2020krp}. Type Ia supernovae can also be used to measure peculiar velocities both at low~\citep{Bonvin:2006en,Appleby:2014kea,Huterer:2015gpa,Castro:2015rrx} and intermediate redshifts~\cite{Howlett:2017asw,Garcia:2020qah}, but the strong inhomogeneity of supernova data poses a challenge. A detailed review of cosmic dipole observables is given by~\cite{Gibelyou:2012ri}.

When dealing with cosmological perturbations one should keep in mind that not all variables are observable, i.e., gauge independent. The definition of the CMB rest frame is likewise not univocal. It can be defined as the frame in which the dipole is zero, or alternatively as the one in which aberration is zero~\cite{Lewis:2006fu,Meerburg:2017xga}, but this choice is unimportant for our analysis. The crucial point is that the degeneracy between primordial and kinematic effects in the CMB dipole can be broken by measuring the Doppler-like and aberration-like couplings in the CMB present in all scales~\cite{Roldan:2016ayx}. In the case of a peculiar velocity these couplings must be present with well-determined coefficients, as discussed in our companion paper \cite{dopp-aber-PRD} (see also~\cite{Notari:2011sb}). In other scenarios, one of both of these couplings can differ~\cite{Roldan:2016ayx}, and therefore by combining the observations of the dipole, Doppler and aberration effects one can independently measure our peculiar velocity and the intrinsic CMB dipole in a more model-independent way. This is what we pursue in the present work.

The aberration and Doppler couplings were shown to be detectable by Planck by~\cite{Kosowsky:2010jm} and~\cite{Amendola:2010ty}, and subsequently measured by~\cite{Aghanim:2013suk} following an estimator proposed in~\cite{Hanson:2009gu}. They made use of only the 143GHz and 217GHz channels and measured $v  \equiv \beta c = 384 \,\rm{km/s} \pm 78 \, \rm{km/s \, (stat)}\pm 115$ km/s (syst). As explained in~\cite{Aghanim:2013suk} this measurement however did not distinguish from possible intrinsic contributions and is at least partly degenerate with the standard dipole measurements. We revisit this issue in more detail in our companion paper~\cite{dopp-aber-PRD}.

In this letter we improve on the results of~\cite{Aghanim:2013suk} in several ways. First of all, we remove the couplings which bring no new information with respect to the dipole. Second, we remove biases in the estimators of aberration and Doppler by simulating the effects using $48^2$ combinations of orientations, which ensures no a priori information on the direction of either effect. Third, we rely on the final Planck component separated maps of SMICA and NILC instead on single frequency maps and use the 2018 data release which is more robust than the 2013 one~\citep{Akrami:2018vks}. Finally, we make use also of the $E$ polarization map. These improvements result not only in better precision and much smaller systematic errors, but allow us to both put an upper bound for the first time on the intrinsic CMB dipole and to make an estimate of our peculiar velocity with respect to the Hubble flow without assuming a negligible intrinsic component.

\textbf{\emph{Estimators and pipeline.}}
The aberration and Doppler estimators and pipelines used here are discussed in detail in our companion paper~\cite{dopp-aber-PRD}, and here we simply summarize the main ideas. The first step is to remove the Dipole Distortions (DD) discussed in~\cite{Notari:2015kla}. The DD produce an extra Doppler signal but are an artifact of the imperfect conversion from intensity into thermodynamic temperature and are degenerate with the dipole. The DD effect depends on the particular weights assigned to each frequency band in a given mapmaking procedure. We therefore estimated the DD for the SMICA and NILC maps produced by the Planck collaboration~\citep{Akrami:2018mcd} in multipole bins of $\Delta \ell = 200$. For SEVEM and Commander the DD signal is highly non-trivial and we do not consider these maps. In the proposed Main Pipeline (MP) the DD are removed at the estimator level, but we also tested a cross-check pipeline (CCP) in which a Doppler boost with the inverse signal is applied to the maps (this also boosts the noise). The pipeline code is available~\cite{aber-dopp:2020}.

We then make use of the ideal estimators proposed in~\cite{Amendola:2010ty}, which ignore biases introduced by the anisotropic noise (present at high $\ell$) and mask. Since both aberration and Doppler result in $\ell, \ell+1$ couplings, there is also an inherit leakage (and thus a correlation) of signal between them. In order to remove these biases and leakages we perform a series of 64 simulations using our Healpix-Boost code \cite{healpix-boost:2020} adding either Doppler or aberration signals in each of 48 different sky directions as well as the anisotropic noise, realistic beaming and mask. For each cartesian component we calibrate the estimators by a $\chi^2$ fit first of 2 nuisance parameters to account for mask and noise, and then 2 additional nuisance parameters which de-correlate both estimators to good precision. This procedure was carried out for both temperature and polarization two point estimators used ($TT$ and $EE$).

This pipeline allows for measurements of aberration ($\bm{\beta}^{\rm A}$) and Doppler ($\bm{\beta}^{\rm D}$) which are independent between themselves and with the dipole. We also consider the more traditional case in which one assumes \emph{a priori} that $\bm{\beta}^{\rm A} \equiv \bm{\beta}^{\rm D} \equiv \bm{\beta}^{\rm B}$ as in a standard boost~(B) transformation (i.e. assuming no additional sources of Doppler-like or aberration-like couplings). This leads to a higher significance detection, but it only serves as a simple cross-check as all physical information is already encoded in the high-precision observation of the temperature dipole. The precision in both $\bm{\beta}^{\rm A}$ and $\bm{\beta}^{\rm D}$ is estimated in~\cite{dopp-aber-PRD}.

\textbf{\emph{Measuring the intrinsic dipole.}}
Besides the standard scenario of a simple Doppler effect on the monopole, a temperature dipole $\Delta_1$ can also be due to an intrinsic dipole component $\Delta_{1,\rm int}$ in the SLS of either the temperature ($\tau$) or gravitational potential ($\phi$) perturbations. A significant contribution from the late Integrated Sachs-Wolfe effect is less likely since the integrand is non-zero only after matter-domination.
On large scales both temperature or gravitational potential perturbations are proportional, but the proportionality depends on whether the perturbations are adiabatic or isocurvature (entropic). For small scales the CMB fluctuations are known to be adiabatic, but for the dipole it could be either or a combination of both.
The nature of the dipole can thus only be understood by adding new observables.

Through a detailed investigation of second-order perturbations on the CMB, the effects of these different primordial scenarios on the Doppler and aberration signals in the CMB was discussed in~\cite{Roldan:2016ayx}. An aberration effect could be mimicked by the dipolar component of gravitational lensing. Lensing is however an integrated quantity and depends on fluctuations along the whole past light-cone. This means that only in fine-tuned models would it produce an aberration signature in the same direction of the dipole and correct amplitude. The Doppler couplings on the other hand are produced by second-order perturbation effects on the SLS. Surprisingly, it was found in~\cite{Roldan:2016ayx} that apart from the standard velocity coupling ($\beta \Theta$, where $\Theta$ represent the linear temperature fluctuations) there are also second-order contributions from $\tau \Theta$ and $\phi \Theta$ which, if Gaussian, produce a total coupling which is given exactly by $\Delta_1 \Theta$ in both adiabatic and isocurvature case. I.e., these couplings are insensitive to the nature of $\Delta_1$. If, on the other hand, the fluctuations contain non-Gaussianities (NG) this degeneracy is broken. In particular, for the much-studied local NG~\cite{Wands:2010af} new terms proportional to $\phi^2$ appear which result in extra $\Delta_1 \phi$ couplings. This changes the amplitude (but not the direction) of the intrinsic Doppler couplings. Thus a detection of conflicting Doppler, aberration and dipole measurements can hint at the presence of a non-standard large intrinsic dipole and allows one to test physics beyond the single-field slow-roll inflation model and modifications on recombination physics~\citep[see, e.g.,][]{Erickcek:2008sm,Lyth:2013vha,Dai:2013kfa,Mathews:2014fwa,Meerburg:2017xga}.

The 3 vector observables are $\bm{\Delta}_{1}$, the aberration $\bm{\beta}^{\rm A}$ and Doppler $\bm{\beta}^{\rm D}$ estimators.  Both $\betaa$ and $\betad$ are the same for $TT$ and $EE$ measurements~\cite{Mukherjee:2013zbi}. Following~\cite{Roldan:2016ayx} and limiting ourselves to local NG we write
\begin{align}\label{eq:intrinsic-obs}
    \bm{\Delta}_1 &= \bm{\beta} + \dint \, , \nonumber \\
    \betad &= \bm{\beta} +  (1+\alpha^{\rm NG}) \dint \,,
    \nonumber\\
    \bm{\beta}^{\rm A} &= \bm{\beta}  + \bm{L}_d \,,
\end{align}
where $\alpha^{\rm NG}=0$  in the absence of NG and $\bm{L}_d$ is the dipolar component of lensing. We assume a constant $\alpha^{\rm NG}$ as a simple parametrization of NG. We thus have 9 observables and 10 unknown quantities in the general case. But for a given $\dint$ model $\alpha^{\rm NG}$ can be computed and one can directly measure $\dint$.
We leave a detailed study on the relation between $\alpha^{\rm NG}$ and the standard NG bispectrum parameter $f_{\rm NL}$ (predicted to be small in the slow-roll inflation scenario~\cite{Maldacena:2002vr,Lyth:2005qj,Creminelli:2004yq}) for future works.

In standard $\Lambda$CDM $|\bm{L}_d|$ can be estimated using linear theory~\cite{Hanson:2009kr}. For Planck's best fit cosmology we find ${L}_d = 2.06\;10^{-4} \simeq 0.17 {\Delta}_1$~\cite{Planck:2018lbu} (see their Fig.~A.1). We include it stochastically in our analysis below, but it represents only a small contribution. We remark that in alternative cosmologies $L_d$ could be larger, and for a more robust measurement of $\dint$ it would be important to independently estimate it in future large-scale structure surveys, as in principle it can be estimated if the intervening matter distribution is known. This interesting subject merits future research.


\setlength\tabcolsep{3pt}
\begin{table}
    \centering
    \begin{tabular}{c l c c c}
    \cmidrule{1-5}\morecmidrules\cmidrule{1-5}
    \multicolumn{2}{c}{{$\boldsymbol{TT}$+$\boldsymbol{EE}$}} & $|\bm{v}|$  [km/s] & $l(^{\circ})$   &   $b(^{\circ})$ \\
    \hline
    & Aberration    & $300\pm 99\pm 13$ & $276\pm 32\pm .1$  & $51\pm 19\pm .7$ \\
    \raisebox{-5.8pt}[0pt][0pt]{\rotatebox[origin=c]{90}{SMICA}} & Doppler       & $390\pm 140\pm 13$ & $210\pm 56\pm 3$  & $-2\pm 30\pm .5$ \\
    & Boost & $321\pm 84\pm 9$ & $234\pm 21\pm .1$  & $43\pm 15\pm .2$ \\
    & Velocity & $300^{+111}_{-93}\pm 13$ & $276\pm 33\pm .1$  & $51\pm 19\pm .7$ \\
    \hline
    & Aberration    & $296\pm 100\pm 10$ & $280\pm 32\pm .3$  & $50\pm 20\pm .3$ \\
    \raisebox{-6.2pt}[0pt][0pt]{\rotatebox[origin=c]{90}{NILC}} & Doppler       & $380\pm 140\pm 10$ & $208\pm 56\pm 2$  & $13\pm 30\pm .2$ \\
    & Boost & $332\pm 83\pm 9$ & $250\pm 22\pm 1$  & $50\pm 15\pm .1$ \\
    & Velocity & $296^{+111}_{-88} \pm 10$ & $280\pm 33\pm .3$  & $50\pm 20\pm .3$ \\
    \cmidrule{1-5}\morecmidrules\cmidrule{1-5}
    \end{tabular}
    \vspace{-0.0cm}
    \caption{Results in galactic coord.~for each estimator and component separation method ($1\sigma$ uncertainties). The first error is statistical; the second is the (subdominant) simulated systematics due to masking, anisotropic noise and leakage between Doppler and aberration. The velocity is estimated from aberration by subtracting the stochastic dipolar lensing.
    }
    \label{tab:final_results}
\end{table}

\textbf{\emph{Results.}} The results for the aberration, Doppler and boost estimators, for both SMICA and NILC for $TT$+$EE$ case are shown in galactic coordinates in Table~\ref{tab:final_results}. For each entry we show our estimates for the statistical and systematic uncertainties. As discussed in~\cite{dopp-aber-PRD} the latter are estimated as the residual discrepancy on a given estimator after correcting for the biases due to masking, noise and leakage of signal between Doppler and aberration.

\begin{figure}
    \centering
    \includegraphics[width=0.87\columnwidth]{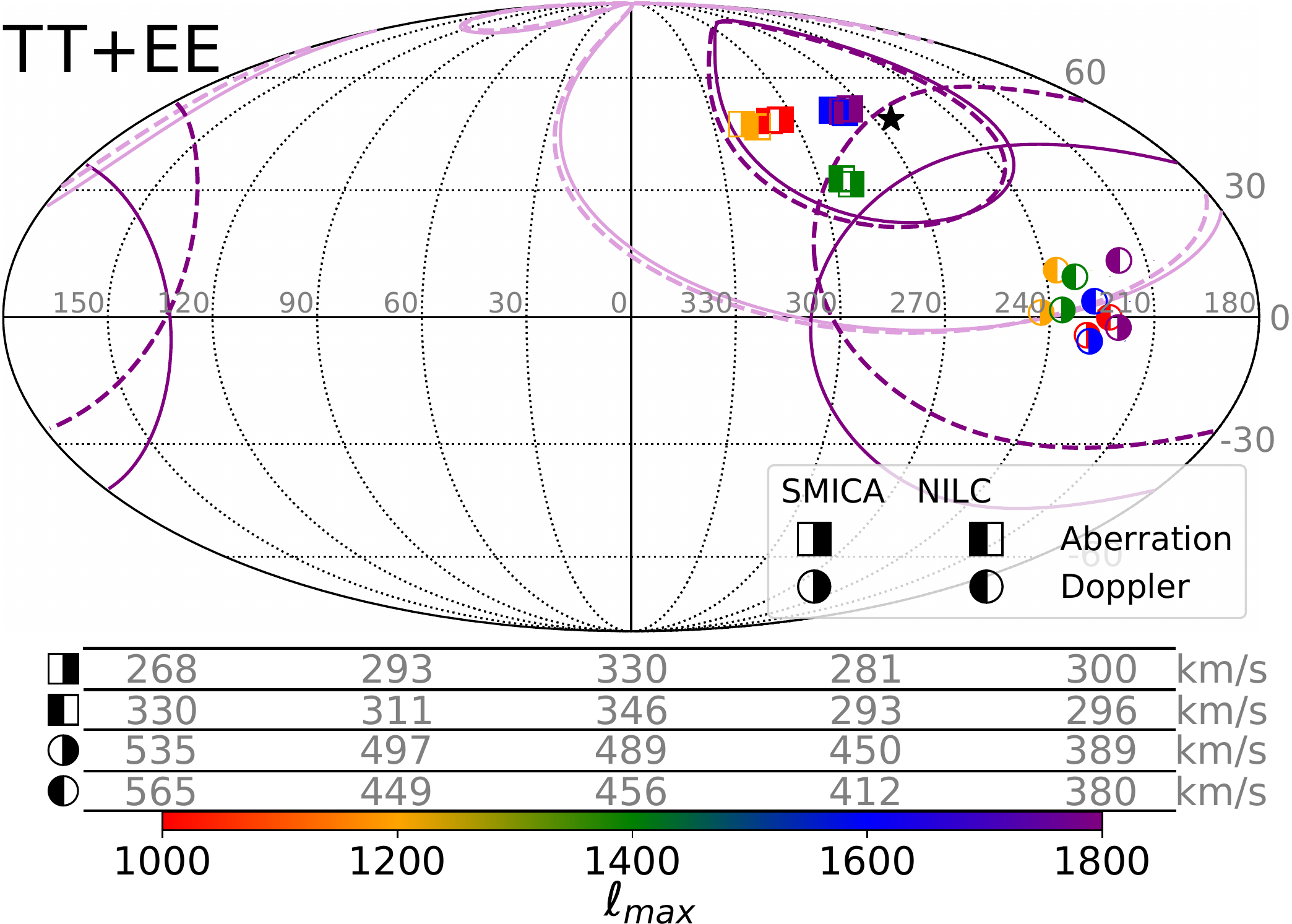}
    \vspace{-0.2cm}
    \caption{Aberration and Doppler estimations as a function of $\ell_{\rm max}$ for the main pipeline for SMICA and NILC. The $\star$ represents the dipole. Solid (dashed) contours are the 1 and 2$\sigma$ confidence levels for SMICA (NILC). The values below the Mollweide plot are the amplitudes of the estimated vectors.
    \label{fig:mollweide_mp_abdopp}
    }
\end{figure}

\begin{figure}
    \centering
    \includegraphics[width=.87\columnwidth]{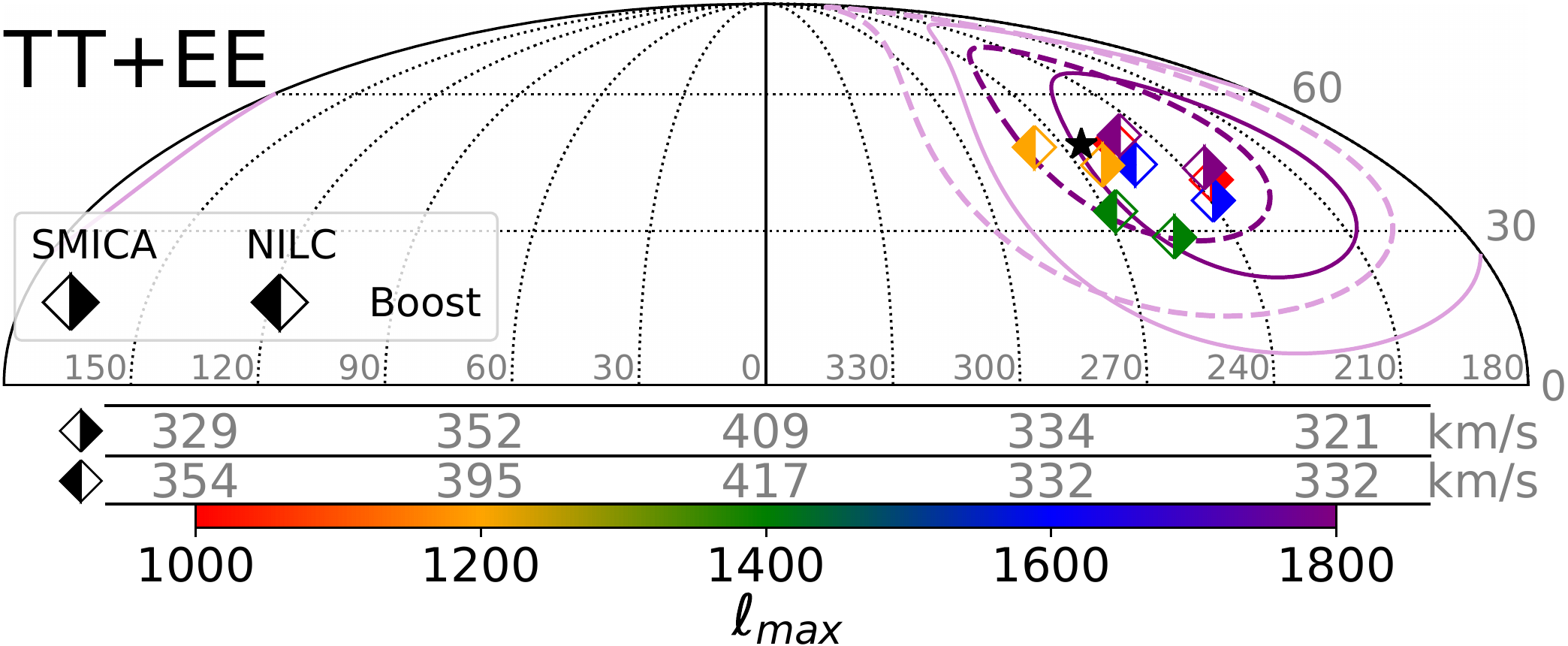}
    \vspace{-0.2cm}
    \caption{Same as Figure~\ref{fig:mollweide_mp_abdopp} for the boost estimator. \vspace{-0.2cm}
    }\label{fig:mollweide_mp_pv}
\end{figure}

\begin{figure}[t!]
    \centering
    \includegraphics[width=0.87\columnwidth]{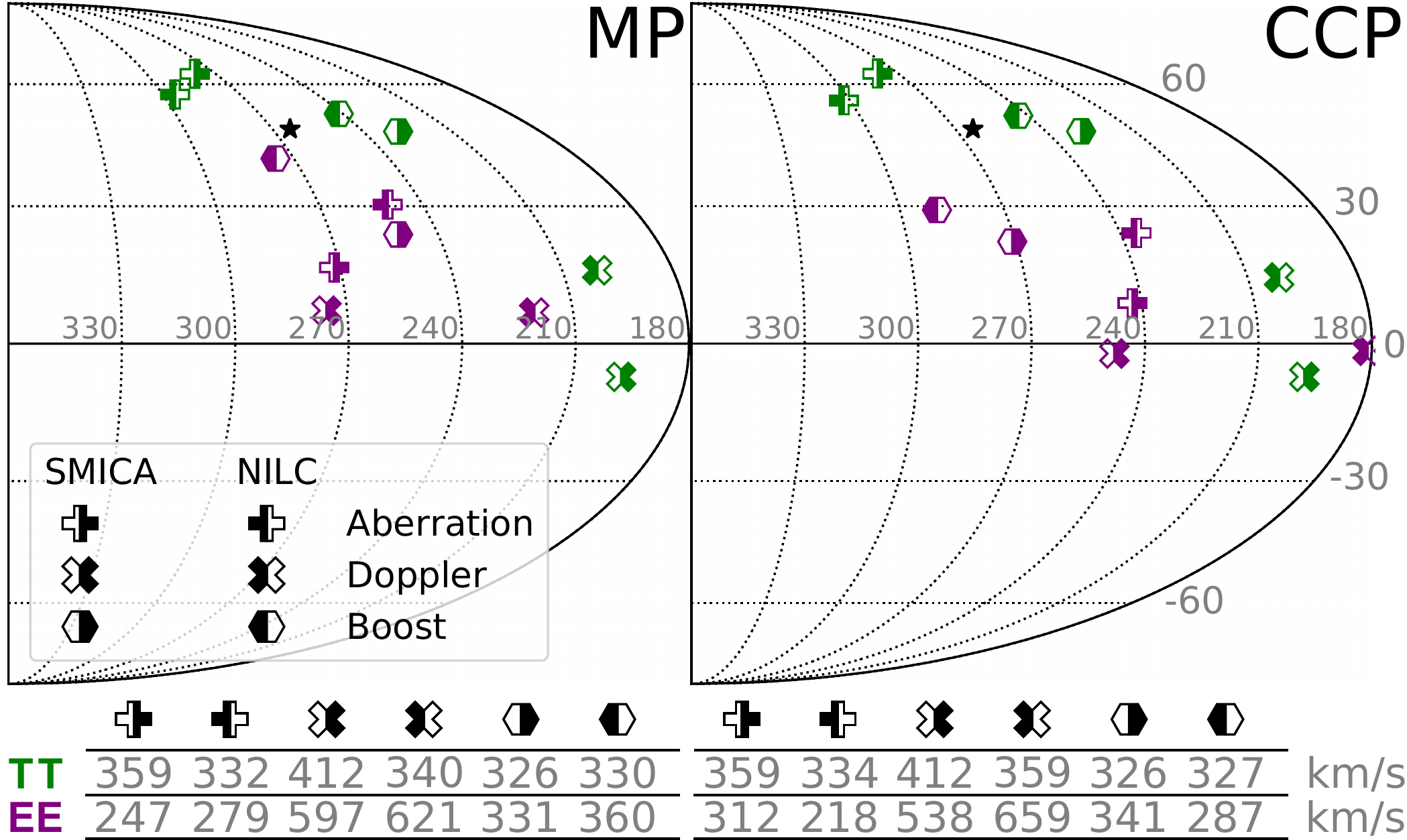}
    \vspace{-0.2cm}
    \caption{Same as Figure~\ref{fig:mollweide_mp_abdopp}, with only eastern hemisphere, for the comparision between MP and CCP for $TT$ and $EE$ measurements with $\ell_{\rm max} = 1800$ and $1150$ respectively.
    }\label{fig:mollweide_mp_vs_ccp}
\end{figure}

Figure~\ref{fig:mollweide_mp_abdopp} shows results for different values of $\ell_{\rm max}$ for $TT$ up to 1800. For $EE$ we stop at 1150. We use always $\ell_{\rm min} = 200$~\cite{dopp-aber-PRD}.  For aberration we draw the 1 and 2$\sigma$ (68.3\% and 95.4\%) confidence intervals (CI) for $TT$+$EE$. The figures use the standard Mollweide projection for galactic coordinates. Figure~\ref{fig:mollweide_mp_pv} shows the boost estimator contours. As expected, since this case assumes the same direction for both aberration and Doppler, this results in greater precision but is only useful as a cross-check. Both MP and CCP pipelines are compared in Figure \ref{fig:mollweide_mp_vs_ccp}.

The statistical significance of each estimator result is quoted in two complementary ways. First, we compare each estimator with the kinematic dipole hypothesis. The second estimate assumes there is no Doppler or aberration effects whatsoever in the data, including Doppler due to the DD, which allows one to get the overall statistical significance of these $\ell, \ell+1$ correlations in the data. Table~\ref{tab:chi2} summarizes the results, which are in agreement with the kinematic hypothesis. Additionally, the correlations are detected at 3.5--3.9$\sigma$ for aberration, 2.6--2.9$\sigma$ for Doppler, and 6.1--6.4$\sigma$ for boost. We also show the combined significance of Doppler and aberration (4.1--4.6$\sigma$), which differs from the boost estimator in not assuming they point in the same direction. For these estimates we rely on the results for the cartesian components instead of the results for the amplitudes and galactic coordinates, as the former require one less bias term and has uncertainties which are Gaussian~\cite{dopp-aber-PRD}. The full ($TT$, $EE$, $TT$+$EE$) cartesian tables  are available~\cite{aber-dopp:2020}.

\setlength\tabcolsep{4pt}
\begin{table}
    \centering
    \begin{tabular}{ c l c c c c }
    \cmidrule{1-6}\morecmidrules\cmidrule{1-6}
    & &  \multicolumn{2}{c}{$\bm{\beta}=\bm{\Delta_1}$} & \multicolumn{2}{c}{$\beta = {\rm DD} = 0 $}  \\
    \cmidrule{1-6}\morecmidrules\cmidrule{1-6}
    $\boldsymbol{TT}$+$\boldsymbol{EE}$   &    &  $\chi^{2}$ & $\sigma$-value  & $\chi^{2}$ & $\sigma$-value  \\ \hline
    & Aberration  \rule{0pt}{3ex}    & 0.3 & 0.1 & 18 & 3.5 \\
    \raisebox{-5pt}[0pt][0pt]{\rotatebox[origin=c]{90}{SMICA}} & Doppler        & 4.6 & 1.3 & 12 & 2.6 \\
    & Boost  & 1.3 & 0.3 & 45 & 6.1 \\
    & Aber. \& Dopp. \rule[-1.4ex]{0pt}{0pt} & 4.9 & 0.6 & 30 & 4.1 \\          \hline
    & Aberration \rule{0pt}{3ex}     & 0.3 & 0.1 & 21 & 3.9 \\
    \raisebox{-6.5pt}[0pt][0pt]{\rotatebox[origin=c]{90}{NILC}} & Doppler        & 2.7 & 0.8 & 13 & 2.9 \\
    & Boost  & 0.4 & 0.1 & 49 & 6.4 \\
    & Aber. \& Dopp. \rule[-1ex]{0pt}{0pt} & 3.0 & 0.2 & 34 & 4.6 \\          \cmidrule{1-6}\morecmidrules\cmidrule{1-6}
    \end{tabular}
    \vspace{-0.2cm}
    \caption{Statistical significance for both component separation methods. The $\beta = \Delta_1$ column assumes the dipole is completely due to our velocity; the  $\beta = {\rm DD} = 0$ column assumes there is no Doppler or aberration effect of any kind.
    }
    \label{tab:chi2}
\end{table}

\begin{figure}
    \includegraphics[trim=0mm 0mm 0mm 0mm, clip, width=0.79\columnwidth]{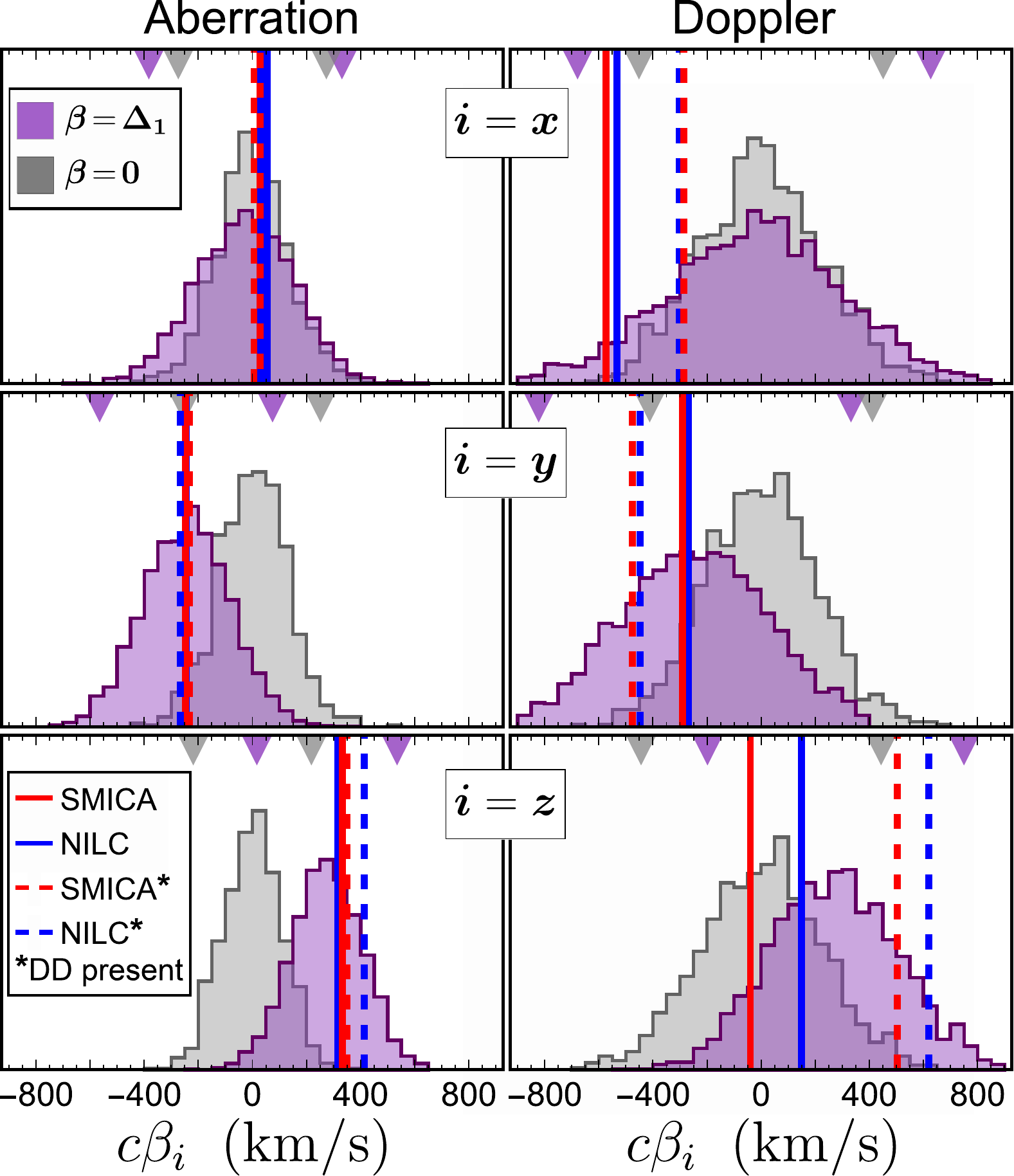}
    \vspace{-0.2cm}
    \caption{Histograms for $\bm{\beta} = \bm{\Delta}_{1}$ (transparent purple) and the null hypothesis $\beta = 0$ (gray) using $TT$+$EE$ centered around the expected values. The measurements are shown as vertical lines for SMICA (red) and NILC (blue); in the solid lines we remove the uninformative DD effect, in the dashed lines it is left included.     The $\pm 2 \sigma$ intervals are depicted by $\blacktriangledown$. Both cases uses $\ell_{\rm max}=1800$ for $TT$ and 1150 for $EE$ and the MP. The solid lines are consistent with the purple diagrams, while the dashed lines and gray histograms yield a large discrepancy.
    \label{fig:hist_delta_beta}}
\end{figure}

The histograms of our aberration and Doppler simulations for each cartesian component are depicted in Figure~\ref{fig:hist_delta_beta}. These results are the base of Table~\ref{tab:chi2}. The purple histograms are the main results, which assume that both Doppler and aberration signals are present (including the DD) and thus one can leak into the estimator of the other. We also center the histograms in the dipole values, to test the kinematic hypothesis. The gray histograms assume no Doppler or aberration effects are present, which means the leakage is not an issue and therefore the statistical errors are smaller, and are used to compute the overall significance of Doppler and aberration in the data.

As discussed above by measuring Doppler and aberration separately we can constrain $\dint$, $\alpha^{\rm NG}$ and $\bm{L}_d$. If we set $\alpha^{\rm NG} \ll 1$, $\betad$ becomes just a cross-check and one can only measure $\done-\betaa = \dint-\bm{L}_d$, which is usually assumed to be zero. In the more general case, $\alpha^{\rm NG} \dint = \betad-\done$ and if we now consider $\bm{L}_d$ stochastically with a known amplitude (see above) but unknown direction one can estimate $\dint$. The uncertainty in $\dint$ is then basically the aberration error (as $\bm{L}_d$ is small). The uncertainty in $\alpha^{\rm NG}$ is more complicated as it follows a ratio distribution, but it is dominated by the Doppler error. The results for $\dint$ and $\alpha^{\rm NG}$ are shown in Table~\ref{tab:intrinsic_stat} using 95\% CI. In all cases the values are consistent with zero, so we only quote limits for $|\dint|$, which turns out to be similar to the observed $\Delta_1$ value.

\setlength\tabcolsep{3pt}
\begin{table}
    \centering
    \begin{tabular}{l c c c c}
    \cmidrule{1-5}\morecmidrules\cmidrule{1-5}
    {{$\boldsymbol{TT}$+$\boldsymbol{EE}$}} & \multicolumn{2}{c}{\qquad\quad$\dint$}  &  \multicolumn{2}{c}{ $\;\alpha^{\rm NG}\;$}   \\
    & amplitude  & $\sigma$-value  & [$95\%$ CI]  & $\sigma$-value \\
    \hline
    SMICA \T  & $<3.6$ mK [95\% CI] & 0.1 & $1.0^{+3.0}_{-3.9}$ & 0.7   \\
    NILC  \T   & $<3.7$ mK [95\% CI] & 0.1 & $0.9^{+2.4}_{-3.3}$ & 0.9 \\
    \cmidrule{1-5}\morecmidrules\cmidrule{1-5}
    \end{tabular}
    \vspace{-0.2cm}
    \caption{
     Measurements and null-hypothesis test of the intrinsic dipole  and its non-Gaussian parameter  using the full $TT$+$EE$ results. The total observed dipole is $3.362$ mK.}
    \label{tab:intrinsic_stat}
\end{table}

We remark that these are the first direct constraints on the intrinsic CMB dipole, the largest mode in the SLS. Our results are fully consistent with the kinematic dipole interpretation and show no sign of non-Gaussianities, as expected by the standard cosmological model. These findings, however, exclude for instance the possibility of a $\sim 1000$~km/s value for $c\beta$, the raw result in many Cosmic Radio dipole measurements~\citep[see e.g.][]{Siewert:2020krp}, which otherwise could be fine-tuned with a large $\dint$ opposite to $\boldsymbol{\beta}$ such that the vector sum resulted in the observed $\done$.


\textbf{\emph{Perspectives.}} In over half a century since the discovery of the CMB we have been able to measure temperature, polarization and lensing to very high $\ell$s. But only now we are  finally able to put physical constraints on the largest possible scale, the temperature dipole.

The measurements of $\boldsymbol{\beta}$ and $\dint$ are limited by the precision of the aberration and Doppler estimators in Planck data. Doppler in particular has high uncertainty: after removing the uninformative DD it is detected at less than $2\sigma$.  Future high-resolution CMB experiments can improve this precision by both probing higher multipoles and by measuring the $E$ modes with higher S/N~\cite{dopp-aber-PRD}.

If $|\dint| \sim 0.03$mK as in the higher multipoles, it will remain undetectable with this method in the foreseeable future, but ruling out exotic scenarios with $|\dint| \sim 1$mK will be feasible in the near future. Although such a large intrinsic dipole would be inconsistent with our standard inflation scenario, there has been numerous claims of large scale anomalies in the CMB~\cite{Schwarz:2015cma}, and so it is important to also try to detect its largest possible scale.

A non-negligible intrinsic dipole would mean that the velocity used to infer cosmological redshifts is not $\simeq 370$ km/s, which can be a source of bias. Supernova surveys are in particular sensitive to accurate redshift determinations~\citep[see e.g.][]{Calcino:2016jpu}. Model-independent measurements of our velocity such as the one from the aberration of the CMB allows for more robust redshift corrections.

Both Doppler and aberration effects should be present in all cosmological observables. The SKA telescope is predicted to measure our velocity in radio continuum maps with 10\% precision~\citep{Bengaly:2018ykb}. Both secular extragalactic parallax measurements using GAIA final release~\citep{Paine:2019vep} and future CMB experiments~\citep{Burigana:2017bxl} are expected to provide a similar precision. Other proposed ways to measure the intrinsic CMB dipole includes the spectral distortions of the monopole and quadrupole in future spectrometric CMB instruments~\citep{Yasini:2016dnd} and the induced effect on the lensing of the CMB for $\ell \gtrsim 3000$~\citep{Meerburg:2017xga}. It remains to be seen which observable will perform best in the future.

\section*{Acknowledgements}
We would like to thank Soumen Basak, Paola Delgado, Luiz Filipe Guimarães, Hans Kristian Eriksen, Maude Le Jeune, Alessio Notari, Omar Roldan, Douglas Scott and Suvodip Mukherjee for useful discussions. We also thank the anonymous referees for helpful suggestions. PSF is supported by the Brazilian research agency CAPES (Coordenação de Aperfeiçoamento de Pessoal de Nível Superior). MQ is supported by the Brazilian research agencies FAPERJ and CNPq (Conselho Nacional de Desenvolvimento Científico e Tecnológico).

\bibliography{cmb}

\label{lastpage}

\end{document}